# Learning Best K analogies from Data Distribution for Case-Based Software Effort Estimation

Mohammad Azzeh, Yousef Elsheikh
Faculty of Information Technology
Applied Science University
Amman, Jordan
e-mail: {m.y.azzeh, y_elsheikh}@asu.edu.jo

*Abstract*— **Case-Based Reasoning (CBR) has been widely used to generate good software effort estimates. The predictive performance of CBR is a dataset dependent and subject to extremely large space of configuration possibilities. Regardless of the type of adaptation technique, deciding on the optimal number of similar cases to be used before applying CBR is a key challenge. In this paper we propose a new technique based on Bisecting k-medoids clustering algorithm to better understanding the structure of a dataset and discovering the optimal cases for each individual project by excluding irrelevant cases. Results obtained showed that understanding of the data characteristic prior prediction stage can help in automatically finding the best number of cases for each test project. Performance figures of the proposed estimation method are better than those of other regular K-based CBR methods.**

## I. INTRODUCTION

Estimating the likely software project effort is a vital task for project planning, control and assigning resources [5, 23, 26, 28]. Although a variety of software effort estimation models have been proposed so far, the Case Based Reasoning (CBR) method is still the most widely investigated method. CBR is a knowledge management method based on premise that history almost repeats itself which leads to problem solving can be based upon retrieval by similarity [30]. It has been favored over regression techniques since software datasets often exhibit complex structure with a lot of discontinuities [2, 6, 20, 21].

The predictive performance of CBR suffers from common problems such as very large performance deviations as well as being highly dataset dependent. This is due to a large space of configuration possibilities and design decisions induced for each individual dataset [16]. Recent publications reported the importance of discovering the optimal K closest cases for generating better estimates in CBR [11, 27]. Conventional K-based CBR methods start with a single analogy and increase this number depending on the overall performance of the whole dataset then it uses the K value that produces the overall best performance. However, a fixed K value that produces overall best performance does not necessarily provide the best performance for individual projects. Our claim is that we can avoid sticking to a fixed best performing number of cases which changes from dataset to dataset or even from a single project to another within the same dataset. We propose an alternative technique to calibrate CBR by using Bisecting k-medoids (BK) clustering algorithm. The k-medoids is a clustering algorithm related to the centroid-based algorithms which groups similar individual instances within a dataset into N clusters known a priori [29, 30]. This enables us to discover the structure of dataset efficiently and automatically come up with the best number of K closest cases as well as excluding irrelevant cases for each individual test instance.

The rest of the article is organized as follows: Section 2 provides the Background of Case-Based Effort Estimation. Section 3 defines the Research question and introduces main problem. Section 4 presents the proposed technique. Section 5 presents experimental design. Section 6 presents the results we obtained. Section 7 presents threats to validity of this study. Lastly, Section 8 summarizes our conclusions and future work.

## II. BACKGROUND

Case-Based effort estimation is a variant of CBR which makes prediction for a new project by retrieving previously completed successful projects that have been encountered and remembered as historical projects [12, 13]. The data driven CBR method involves four major stages [25]: (1) retrieve the most similar training projects using Euclidean distance function as depicted in Eq. 1. Then (2) reuse the past solutions from the set of retrieved analogues to solve the new problem. (3) revise the proposed solution and to better adapt the target problem. Finally, (4) retain the solved problem for future problem solving.

$$d(p_i, p_j) = \frac{1}{m}\sqrt{\sum\nolimits_{t=1}^{m} \Delta(p_{it}, p_{jt})} \qquad (1)$$

where $d$ is the similarity measure. $m$ is the number of predictor features, $t$ is the index of feature, $p_i$ and $p_j$ are projects under investigation and:

$$\Delta(p_{it}, p_{jt}) = \begin{cases} \dfrac{(p_{it} - p_{jt})^2}{\max_t - \min_t} & \text{if } t \text{ is continuous} \\ 0 & \text{if } t \text{ is categorical and } p_{it} = p_{jt} \\ 1 & \text{if } t \text{ is categorical and } p_{it} \neq p_{jt} \end{cases} \qquad (2)$$

Although CBR generates successful performance figures in certain datasets, it still suffers from local tuning problems when they were to be applied in another setting [3]. Local tuning requires mainly learning appropriate K cases that fits procedure of adjustment and reflects dataset characteristics [16]. The classical approach uses a fixed number of cases (K=1, or 2 or…etc.) for all test projects, which is somewhat considered simpler but it relies heavily on the estimator intuitions [3]. In this direction, Kirsopp et al. [15] proposed making predictions from the K=2 nearest cases as it was found as the best value for their investigated datasets. In a further study Kirsopp et al. have increased their accuracy values with case and feature subset selection strategies. On the other hand, Idri et al. [9] proposed using all projects that fall within a certain similarity threshold. This approach could ignore some useful projects which might contribute better when similarity between selected and unselected cases is negligible. Li et al. [17] performed rigorous trials on actual and artificial datasets and they observed effect of various K values. However, we believe that reflection on dataset prior applying to different algorithms under multiple settings is of more significance. But, this is not enough because the selection of K cases is not only a dataset dependent but also adjustment method dependent. In this study we focus only on discovering the best number of cases to be used for each individual test project from the characteristics of the dataset.

## III. RESEARCH QUESTIONS

Finding the appropriate number of cases to be used in CBR is a challenge on its own, and has a strong impact on the overall predictive performance. Conventional K-based CBR methods start with a single analogy and increase this number depending on the overall performance of the whole dataset then it uses the K value that produces the overall best performance. However, a fixed K value that produces overall best performance does not necessarily provide the best performance for individual projects. Furthermore, previous studies reported that the proper selection of K cases is a dataset dependent and subject to underlying distribution of the dataset [16]. For these reasons, we propose a new technique based on Bisecting k-medoids clustering algorithm to find the optimal number of cases to tune and configure CBR method. To the best of our knowledge, it has not been used previously in software effort estimation domain. Unlike regular K-based CBR methods, the proposed technique starts with all projects in the train dataset and gradually excludes irrelevant projects on the basis of compactness degree. The proposed work attempts to answer the following research questions:
1. How can we better understand the characteristics of a particular dataset and dynamically come up with optimum K number of analogies?
2. Does the performance of CBR improve with automatic dynamic selection of K cases for each individual project?

## IV. THE PROPOSED CBR BASED BISECTING K-MEDOIDS ALGORITHM CBR(BK)

The k-medoids is a clustering algorithm related to the centroid-based algorithms which groups similar individual instances within a dataset into N clusters known a priori [29, 30]. It is more robust to noise and outliers as compared to k-means because it minimizes the sum of pairwise dissimilarities instead of a sum of squared Euclidean distances. A medoid can be defined as the instance of a cluster, whose average dissimilarity to all the instances in the cluster is minimal i.e. it is a most centrally located point in the cluster. The popularity of making use of k-medoids clustering is its ability to use arbitrary dissimilarity or distances functions, which also makes it an appealing choice of clustering method for software effort data as software effort datasets also exhibit very dissimilar characteristics.

Regardless of the k-medoids algorithm advantages it still has some challenges such as guessing the number of clusters that can be used to find the partitions that best fits the underlying data [30]. To avoid this challenge we employed bisecting procedure with k-medoids algorithm and propose Bisecting k-medoids algorithm (BK). BK is a variant of k-medoids algorithm that can produce hierarchical clustering by recursively applying the basic k-medoids. It starts by considering the whole dataset to be one cluster. At each step, one cluster is selected and bisected further into two sub clusters using the basic k-medoids. Note that by recursively using a bisecting k-medoids clustering procedure, the dataset can be partitioned into any given number of clusters in which the so-obtained clusters are structured as a hierarchical binary tree. The decision whether to continue clustering or stop it depends on the comparison of compactness degree between childes and their direct parent in the tree. If the maximum of compactness of child clusters is smaller than compactness of their direct parent then clustering is continued. Otherwise it is stopped and the parent cluster is considered as a leaf node. This criterion enables the BK to uniformly partition the dataset into homogenous clusters. In this paper the average cluster compactness as a measure of homogeneity of each cluster is used, it is defined as:

$$Compactness = \frac{1}{n}\sum_{i=1}^{k}\sum_{j=1, x_j \in C_i}^{n} \|x_j - v_i\|^2 \quad (3)$$

where $\|\cdot\|$ is the usual Euclidean norm, $x_j$ is the $j^{th}$ data object, $v_i$ is the center of $i^{th}$ cluster ($C_i$) and $k$ is the number of clusters. A smaller value of this measure indicates a high homogeneity (less scattering).

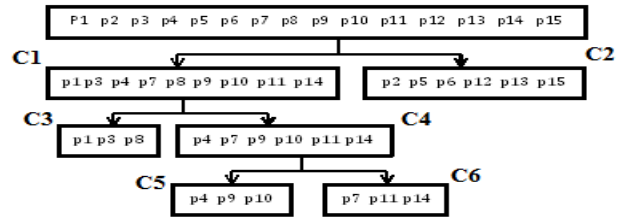

Figure 1. Illustration of Bisecting k-medoids algorithm

Figure 1 is a very simple BK tree that can be formed on a simple dataset of 15 projects. It shows how the main cluster is bisected recursively into four leaf clusters are: C2, C3, C5 and C6. Tricky point here is that unlike K-based CBR methods, BK does not need any expert interference to discover dataset characteristics so as to decide on the number of trees to be built or the number of cases to be used in estimation. To better understand the BK algorithm, see the pseudo code in Figure 2.

*1: **Input**: The dataset X*
*2: **Output**: The set of k clusters S={C1, C2, C3, C4, ..Ck}*
*3: **Initialization**: Let V=X , S={}, NextLevl={}*
*4: **Repeat** while size(V)> 0*
*5: **foreach** Cluster C in V*
*6:       Comp ← compactness (C)*
*7:       [C1,C2] ← k-medoids(C,2)*
*8;       Comp1 ← compactness(C1)*
*9:       Comp2 ← compactness(C2)*
*10:      If(max(Comp1,Comp2)<Comp)*
*11:          NextLevel ← NextLevel ∪ {C1,C2}*
*12:      Else*
*13:          S ← S ∪ {C}*
*14:   **End***
*15:   V ← NextLevel*
*16:   NextLevel ← {}*
*17: **End***

Figure 2. Bisecting k-medoids algorithm

Finally once BK tree is built, the estimation process starts. The un-weighted mean effort of the train projects of the leaf cluster whose medoid is closest to the test project becomes the estimated effort value for that test project as shown in Eq. 4. i.e. we choose to use K-many cases for estimation where K is the number of train instances that are in the selected cluster.

$$Effort(p_t) = \frac{1}{K} \sum_{i=1}^{K} Effort(p_i) \qquad (4)$$

## V. EXPERIMENTAL DESIGN

As it was reported [16] most of the methods in literature were tested on a single or a very limited number datasets, thereby reducing the credibility of the proposed method. To avoid this pitfall, we included nine datasets from two different sources namely PROMISE [4] and ISBSG [10]. PROMISE is an on-line publically available data repository and it consists of datasets donated by various researchers around the world. The datasets come from this source are: Desharnais [7], Kemerer [14], Albrecht [1], COCOMO [4], Maxwell [19], Telecom [4] and NASA93 [4] datasets. The other dataset comes from ISBSG data repository (release 10) which is a large data repository consists more than 4000 projects collected from different types of projects around the world. Since many projects have missing values only 500 projects with quality rating "A" are considered. 14 useful features were selected, 8 of which are numerical features and 6 of which are categorical features. The descriptive statistics of such datasets are summarized in Table 1.

TABLE 1 Statistical properties of the datasets

| Dataset | Cases # | Effort min | Effort max | Effort mean |
|---|---|---|---|---|
| ISBSG | 500 | 668 | 14938 | 2828.5 |
| Desharnais | 77 | 546 | 23940 | 5046.3 |
| COCOMO | 63 | 5.9 | 11400 | 683.5 |
| Kemerer | 15 | 23.2 | 1107.3 | 219.2 |
| Albrecht | 24 | 0.5 | 105.2 | 21.87 |
| Maxwell | 62 | 583 | 63694 | 8223.2 |
| NASA93 | 18 | 8.4 | 824 | 624.4 |
| China | 499 | 26 | 54620 | 3921 |
| Telecom | 18 | 23.45 | 1115.5 | 284.3 |

For each dataset we follow the same testing strategy, we used Leave-one-out cross validation to identify test and train projects such that, in each run, we select one project as test set and the remaining projects as training set. This procedure is performed until all projects within dataset are used as test projects. In each run, The prediction accuracy of different techniques is assessed using MMRE, PRED(0.25) performance measure as shown in Eqs. 5 and 6. MMRE computes mean of the absolute percentage of error between actual and predicted project effort values. PRED(0.25) is used as a complementary criterion to count the percentage of estimates that fall within less than 0.25 of the actual values.

$$MMRE = \sum_{i=1}^{N} \frac{|Effort(p_i) - \overline{Effort(p_i)}|}{Effort(p_i)} \qquad (5)$$

where $Effort(p_i)$ and $\overline{Effort(p_i)}$ are the actual value and predicted values of project $p_i$.

$$PRED(0.25) = \frac{\lambda}{N} \times 100 \qquad (6)$$

where $\lambda$ is the number of projects that have magnitude relative error less than 0.25, and $N$ is the number of all observations. We also used Wilcoxon sum rank test to investigate the statistical significance of all the results, setting the confidence limit at 0.05. The Wilcoxon sum rank test is a nonparametric test that compares the medians of two samples. The reason behind using these tests is because all absolute residuals for all models used in this study were not normally distributed. In turn, the obtained results from the proposed approach have benchmarked to other regular K-based CBR methods that use a fixed number of $K$ cases.

## VI. RESULTS

### A. Results for Research Question 1

This study explores the feasibility of learning best K analogy number from the dataset structure prior building

CBR method. The previous results and conclusions indicate that a single best performing K value that is producing the lowest MRE values for the whole dataset does not necessarily produce lowest MRE value for every single project. To illustrates our viewpoint and better understand this problem we carried out an extensive search to find the mean effort value of the best K number of analogies that produces lowest MRE value for every single test project as shown in Figure 3. For a dataset of size n, the best K value can range from 1 to n − 1. Since a few number of datasets were enough to illustrate our viewpoint, we selected 3 datasets that vary in the size (i.e. one small dataset (Albrecht), one medium (Maxwell) and one large (Desharnais)). Figure 3 shows the histogram of best selected K numbers for the three examined datasets, where x-axis represents K analogy number and y-axis represents frequency of K number (i.e. number of projects that chose that K value). It is clear that there is no global K number for all projects, for example in Albrecht dataset, two projects selected only the closest case (K=1), whilst four projects selected two closest cases (K=2) and so on. This indicates that every single project favors different number of closest analogies. The conclusion can be drawn here that using a fixed K number of cases for all test projects will far from optimum and there is provisional evidence that choosing of best K analogy for each individual project is relatively subject to data structure.

The conclusion can be drawn from previous empirical results that the optimum K number of analogies is not global and every project favors different K number. However, our first research question was how can we better understand the characteristics of a particular dataset and dynamically come up with optimum K number of analogies? In this paper we proposed Bisecting k-medoids algorithm to better understand the characteristics of software datasets and automatically come up with the optimum K number. To illustrate that, we executed CBR(BK) over all employed datasets and we recorded the best obtained K for every test project. Figure 4 shows the histogram of K number of analogies for every test project. This demonstrates the capability of BK technique to dynamically discovering the various K values for every test project that takes into account the characteristics of each dataset on the basis of compactness degree. The procedure of selecting has become easier than first (i.e. where the estimator intuition is heavily used to choose the optimum number of analogy) since the entire best K selection process has been left to the BK. The performance figures of the proposed technique are discussed in the next section.

### B. Results for Research Question 2

The second research question was whether the predictive performance of CBR method can be improved when using BK algorithm? Apart from being able to choose the number of cases for each test instance on its own, BK outperforms all the other K-based CBR methods as can be seen in Table 2.

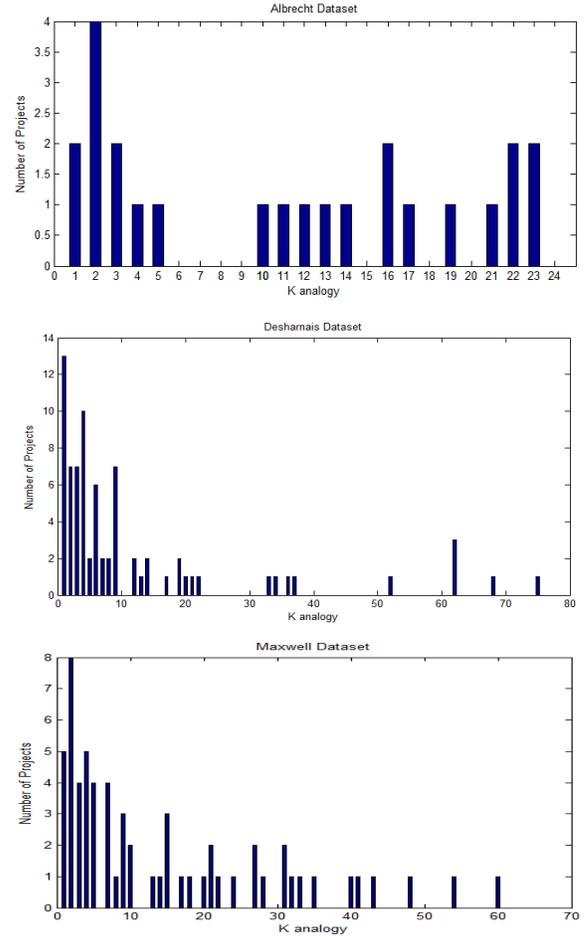

Figure 3. Distribution of K-cases

TABLE 2 MMRE results

| Dataset | CBR (BK) | CBR (K=1) | CBR (K=2) | CBR (K=4) | CBR (K=8) | CBR (K=16) |
|---|---|---|---|---|---|---|
| Albrecht | **45.4** | 71.0 | 66.5 | 73.9 | 89.1 | 146.5 |
| Kemerer | **41.9** | 55.9 | 77.7 | 86.2 | 91.5 | N/A |
| Desharnais | **29.4** | 60.2 | 51.5 | 50.2 | 61.0 | 79.9 |
| COCOMO | **60.43** | 157.1 | 363.2 | 327.3 | 401.8 | 606.55 |
| Maxwell | **41.3** | 182.6 | 132.7 | 149.3 | 138.2 | 145.6 |
| China | **27.7** | 45.2 | 44.2 | 48.5 | 53.8 | 63.0 |
| Telecom | **35.7** | 60.0 | 45.2 | 77.4 | 115.3 | 175.3 |
| ISBSG | **37.0** | 72.6 | 73.2 | 74.7 | 71.7 | 71.7 |
| NASA | **39.4** | 81.2 | 97.5 | 77.6 | 77.1 | 227.6 |

When we look closer at the MMRE values in Table 2, we can see that in all 9 datasets, BK has never been outperformed by other methods with the lowest MMRE values, which suggest that BK has attained better predictive performance values than all other regular K-based CBR methods. This also shows the capability of BK to support small-size datasets such as in Kemerer and Albrecht. However, although it proved inaccurate in this study, the strategy of using fixed K-analogy the effort values may be appropriate in situations where a potential analogues and target project are similar in size feature and other effort drivers. On the other hand, There may be little basis for believing that either increasing or decreasing the K-cases effort values of K-based CBR methods will not improve the accuracy of the estimation.

Table 3 shows that the proposed technique has achieved larger PRED values over eight datasets, which demonstrated that most of the predictions have very good accuracy with MRE vales are less than 0.25. However, overall results from Tables 2 and 3 revealed that there is reasonable believe that using dynamic K-cases for every test project has potential to improve prediction accuracy of CBR in terms of PRED. Concerning discontinuities in the dataset structure, there is clear evidence that the proposed method has capability to group similar projects together in the same cluster as appeared in the results of Maxwell, COCOMO, Kemerer and ISBSG.

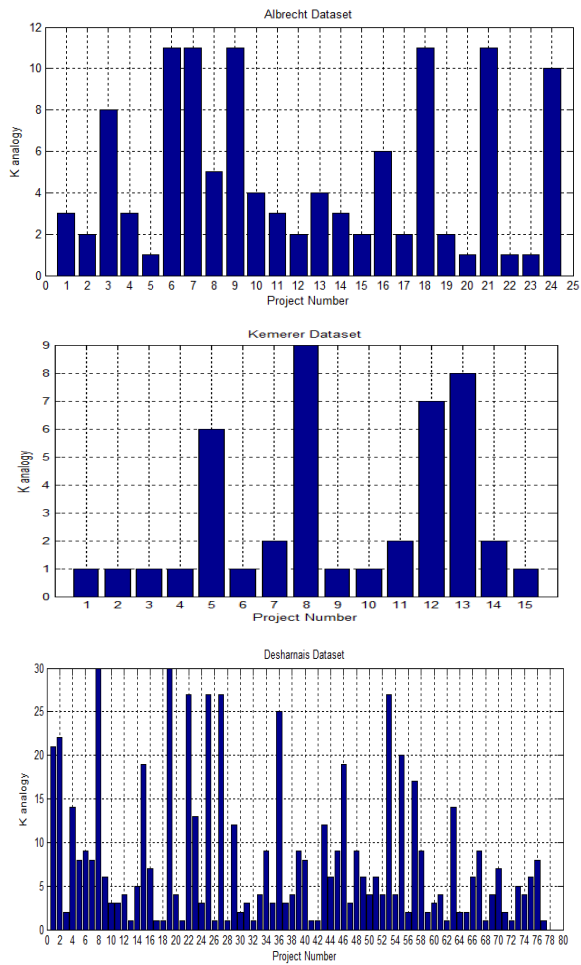

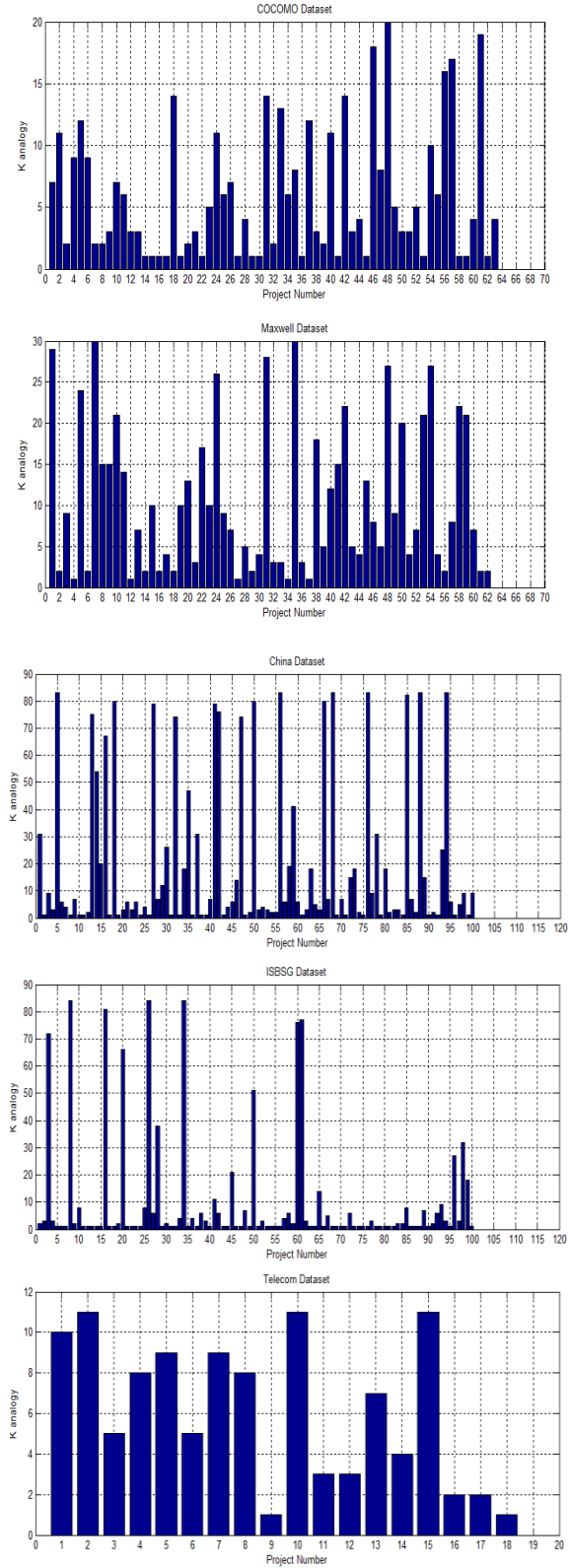

Figure 4. Histogram of K analogies obtained by CBR(BK) for all employed datasets

TABLE 3 PRED results

| Dataset | CBR (BK) | CBR (K=1) | CBR (K=2) | CBR (K=4) | CBR (K=8) | CBR (K=16) |
|---|---|---|---|---|---|---|
| Albrecht | **40.8** | 29.2 | 33.3 | 37.5 | 37.5 | 33.3 |
| Kemerer | **43.3** | 40.0 | 20.0 | 13.3 | 20 | N/A |
| Desharnais | **40.3** | 31.2 | 31.2 | 37.6 | 32.5 | 22.1 |
| COCOMO | **19.3** | 12.7 | 19.1 | 15.9 | 12.7 | 12.7 |
| Maxwell | **22.6** | 9.7 | 19.4 | 14.5 | 16.1 | 29 |
| China | **52.7** | 38.3 | 43.5 | 41.9 | 38.1 | 33.7 |
| Telecom | **53.3** | 33.3 | 50 | 44.4 | 38.9 | 22.2 |
| ISBSG | 38.4 | **39.6** | 30.7 | 29.7 | 25.7 | 22.2 |
| NASA | **39.3** | 33.3 | 38.9 | 22.2 | 11.1 | 0 |

The variants of CBR methods are taken and compared using Wilcoxon sum rank test. The results of Wilcoxon sum rank test of absolute residuals are presented in Table 4. Surprisingly, predictions based on CBR(BK) model presented statistically significant but necessarily accurate estimations than others, confirmed by the results of MMRE as shown in Table 2. Except for small datasets such as Albrecht, Kemerer, Telecom and NASA, the statistical test results demonstrate that there is no significant difference if the predictions generated by any CBR(BK) and other regular *K*-based CBR methods. So it seems that the small datasets are the most challenging ones. These datasets have relatively small number of instances and large degree of heterogeneity between projects so it is difficult to obtain a cluster of sufficient number of instances.

TABLE 4 Wilcoxon sum rank test results

| Dataset | CBR (K=1) | CBR (K=2) | CBR (K=4) | CBR (K=8) | CBR (K=16) |
|---|---|---|---|---|---|
| Albrecht | 0.8 | 0.64 | 0.39 | 0.45 | 0.69 |
| Kemerer | 0.7 | 0.17 | 0.07 | 0.07 | 0.69 |
| Desharnais | **0.01*** | **0.01*** | **0.03*** | **0.01*** | **0.01*** |
| COCOMO | **0.03*** | **0.01*** | **0.02*** | **0.03*** | **0.01*** |
| Maxwell | **0.01*** | **0.01*** | **0.01*** | **0.01*** | **0.04*** |
| China | **0.01*** | **0.01*** | **0.01*** | **0.01*** | **0.01*** |
| Telecom | 0.79 | 0.76 | 0.42 | 0.91 | **0.03*** |
| ISBSG | **0.04*** | **0.01*** | **0.01*** | **0.01*** | **0.01*** |
| NASA | 0.68 | 0.84 | 0.19 | **0.04*** | **0.01*** |

## VII. THREAT TO VALIDITY

This section presents the comments on threats to validities of our study based on internal, external and construct validity. Internal validity is the degree to which conclusions can be drawn with regard to configuration setup of BK algorithm including: 1) the identification of initial medoids of BK for each dataset, 2) determining stopping criterion. Currently, there is no efficient method to choose initial medoids so we used random selection procedure. So we believe that this decision was reasonable even though it makes the k-medoids is computationally intensive. For stopping criterion we preferred to use the compactness performance measure to see when the BK should stop. Although there are plenty of compactness measures we believe that the used measure is sufficient to give us indication of how instances in the same clusters are strongly related.

Concerning construct validity which assures that we are measuring what we actually intended to measure. However, despite special emphasis was placed on the effectiveness of the performance measures, complete certainty with regard to this issue was challenged and we had to rely on common estimation-error based performance measures such as MMRE and PRED, which we no longer believe to be a completely trustworthy accuracy indicator [8, 24]. We do not consider that choice was a problem because (1) They are practical options for majority of researchers [2, 11, 13, 16, 22], and (2) using such measures enables our study to be benchmarked with previous effort estimation studies. On the other hand, in order to make apple-to-apple comparisons between different adaptation techniques we preferred to use Leave-one cross-validation strategy, though some authors favored n-Fold cross validation. The principal reason is that, the Leave-one cross-validation has been used in some previous studies and recommended to do comparison between different estimation models.

With regard to external validity, i.e. the ability to generalize the obtained findings of our comparative studies, we used 8 datasets from 2 different sources to ensure the generalizability of the obtained results. The employed datasets contain a wide diversity of projects in terms of their sources, their domains and the time period they were developed in. We also believe that reproducibility of results is an important factor for external validity. Therefore, we have purposely selected publicly available datasets. However, we consider that some datasets are very old to be used in software cost estimation because they represent different software development approaches and technologies. The reason for this is that these datasets are publically available, and still widely used for benchmarking purposes.

## VIII. CONCLUSION AND FUTURE WORK

This paper proposed a new technique based on utilizing Bisecting k-medoids clustering algorithm and compactness degree to find the best K analogies number from the structure of dataset for each test project. Thus, rather than proposing a fixed best-K value a priori as the traditional CBR methods do, what CBR(BK) does is starting with all the training samples in the dataset, learning the dataset to form BK binary tree and excluding the irrelevant cases on the basis of compactness degree and then discovering the best-K value for each individual project. The proposed technique has the capability to support different-size datasets that have a lot of categorical features. Empirical results on various datasets indicate the performance of the proposed method over other regular K-based CBR methods. So the conclusion can be drawn that the choice of best K value is subject to the characteristics of a software dataset, and this value should be discovered from the structure of dataset. A future work is planned to study the impact of feature selection and weighting on discovering the optimal K value.

ACKNOWLEDGMENT

The authors are grateful to the Applied Science University, Amman, Jordan, for the financial support granted to cover the publication fee of this research article.